\documentclass[twocolumn,aps,showpacs,superscriptaddress,prb,tightenlines,amsmath,amssymb]{revtex4-1}
\usepackage{comment}
\usepackage[colorlinks, linkcolor=blue, anchorcolor=blue, citecolor=blue]{hyperref}

\usepackage{graphicx}
\usepackage{amssymb}
\usepackage{dcolumn}
\usepackage{amsmath}
\usepackage{physics}
\usepackage{mathrsfs}  
\usepackage{color}
\usepackage{bm}
\usepackage{colordvi}

\makeatletter

\newcommand{\Rmnum}[1]{\expandafter\@slowromancap\romannumeral #1@}
\makeatother

\begin{document}

\title{Collective relaxation eigenmodes and anisotropic magnon thermal transport in $\alpha$-MnTe}

\author{Rui-Jie Yao}
\affiliation{Department of Physics, YanTai University, Yantai, Shandong, 264005, People's Republic of China}
\author{Lei Wang}
\email{wang_lei@btbu.edu.cn}
\affiliation{Department of Physics, Beijing Technology and Business University, Beijing 100048, People‘s Republic of China  }
\author{Bo-Ye Sun}
\email{boyesun@ytu.edu.cn}
\affiliation{Department of Physics, YanTai University, Yantai, Shandong, 264005, People's Republic of China}

\date{\today}

\begin{abstract}
The intrinsic magnon thermal transport in the altermagnet $\alpha$-MnTe is studied by solving the three-dimensional linearized Boltzmann transport equation with the four-magnon collision matrix. Diagonalizing the collision matrix gives direct access to the relaxation eigenmodes beyond the relaxation-time approximation. We show that Umklapp scattering lifts the momentum-related zero modes of the Normal-only collision operator and substantially modifies the low-lying relaxation spectrum. Using the same collision matrix, we compute the magnon thermal conductivity tensor. The full linearized Boltzmann transport equation result exceeds the relaxation-time approximation by more than an order of magnitude at low temperature and reveals a strong transport anisotropy, with the out-of-plane thermal conductivity remaining larger than the in-plane component over the studied temperature range. A mode-resolved analysis shows that the dominant heat-carrying modes retain momentum-like character inherited from the Normal-only zero modes, and that the larger out-of-plane conductivity mainly originates from the stronger out-of-plane group-velocity contribution, rather than from a large difference in relaxation lifetimes.
\end{abstract}

\maketitle

\section{\label{sec:introl}Introduction}
Traditional spintronics heavily relies on the spin degree of freedom of electrons, which inherently faces the unavoidable bottleneck of Joule heating limits\cite{chumak2015magnon,safranski2017spin,li2018roles}. To overcome this fundamental barrier, magnetic insulators have emerged as a compelling alternative, where spin and heat transport are mediated by magnons, which are the elementary excitations of magnetic order\cite{philipp2021advances,hou2019spin}. Unlike charge-based electronics, magnon transport operates without Ohmic losses, offering intrinsic advantages such as extremely low damping, tunable frequencies up to the terahertz (THz) regime and rich nonlinear interactions, providing a new paradigm for efficient low-power information processing\cite{check2023chiral,kruglyak2010magnonics,bauer2012spin,barman_2021the,jungwirth2016antiferromagnetic,chumak2022advances,baltz2018antiferromagnetic}.

To further advance magnonic technologies, it is important to explore magnetic properties and spin excitations in emerging magnetic materials. Recently, altermagnetism has emerged as a distinct class of magnetic order that bridges key features of conventional ferromagnetism and antiferromagnetism\cite{smejkal2022emerging, smejkal2022beyond}. Similar to antiferromagnetism, altermagnets possess zero net macroscopic magnetization, which provides robustness against stray magnetic fields and supports ultrafast THz spin dynamics\cite{check2023chiral}. At the same time, their symmetry-enabled anisotropic magnon band structures can generate strongly direction-dependent magnon group velocities and scattering phase space. As a result, altermagnets may exhibit anisotropic thermal and spin transport, allowing direction-selective control of magnonic heat currents. These features make altermagnets a promising platform for high-density memory, ultrafast spintronic devices, THz nanotechnology, and the exploration of unconventional magnon transport phenomena\cite{smejkal2022giant, mazin2022altermagnetism, bai2022observation,smejkal2020crystal,smejkal2022giant}.

Despite rapid progress in altermagnetic and low-dimensional magnetic systems, a detailed microscopic understanding of magnon transport remains limited. In particular, although magnon-magnon scattering processes have been extensively investigated, their connection to macroscopic heat conduction and intrinsic relaxation dynamics is still not fully understood\cite{dyson1956thermodynamic,dyson1956general,sourounis2024impact,garanin1992the,holstein1940field,michael1969effects}. Many theoretical studies of magnon spin and heat transport use a single-mode relaxation-time approximation (RTA) for simplicity \cite{cornelissen2016magnon,schmidt2018boltzmann,streib2019magnon,nakata2021magnonic,arias2025theory,parodi2026thermal,liu2019collective}.
In this approximation, the full linearized collision integral is replaced by independent relaxation channels characterized by mode-dependent relaxation times. While useful, this single-mode picture neglects the intermode coupling encoded in the off-diagonal structure of the collision operator, which can mix different momentum states into collective relaxation eigenmodes. Resolving these eigenmodes requires solving the linearized Boltzmann transport equation (LBTE) with the full microscopic collision matrix, which also provides a direct way to separate the roles of momentum-conserving Normal scattering and momentum-relaxing Umklapp scattering in intrinsic magnon transport\cite{cepellotti2015phonon}.

A concrete material platform for addressing these issues is the 
$g$-wave altermagnet $\alpha$-MnTe\cite{krempasky2024altermagnetic}. 
This compound crystallizes in the hexagonal NiAs-type structure and orders magnetically below a Néel temperature of about $307\,{\rm K}$\cite{kriegner2017magnetic}. Its magnetic structure is characterized by ferromagnetically aligned Mn within the basal planes and antiferromagnetic coupling between adjacent planes, leading to a compensated magnetic state with vanishing net magnetization. Importantly, the exchange interactions in $\alpha$-MnTe are strongly anisotropic: the dominant interlayer antiferromagnetic exchange is much larger than the in-plane ferromagnetic couplings, as established by inelastic neutron scattering measurements\cite{szuszkiewicz2006Spin-wave}. This exchange hierarchy directly shapes the spin-wave dispersion and suggests that both magnon velocities and scattering phase space can be highly direction dependent. In addition, in the undoped limit, the semiconducting character of
$\alpha$-MnTe suppresses electronic heat transport compared with metallic
magnets, allowing the intrinsic magnon contribution to be analyzed more
cleanly at the theoretical level\cite{mu2019phonons}.

These features make $\alpha$-MnTe a useful platform for examining how direction-dependent velocities and collision processes jointly determine magnon heat transport. In the full LBTE, the collision matrix determines collective relaxation eigenmodes and their relaxation rates, whereas the directional group velocities enter through the heat-current driving term. This makes it possible to distinguish whether the anisotropic conductivity is mainly controlled by different collective relaxation rates, or by the stronger velocity-weighted driving of selected relaxation modes.

In this work, we present a microscopic study of intrinsic magnon thermal transport in $\alpha$-MnTe by constructing and diagonalizing the four-magnon collision matrix of the three-dimensional LBTE. This eigenmode-based approach gives direct access to the low-lying relaxation spectrum and allows us to resolve how different relaxation eigenmodes contribute to heat transport. By comparing collision operators with and without Umklapp processes, we clarify how momentum-conserving Normal scattering and momentum-relaxing Umklapp scattering shape the relaxation spectrum, including the lifting of momentum-related zero modes and the mixing of low-lying relaxation eigenmodes.

Using the same collision matrix, we compute the magnon thermal conductivity tensor and benchmark the full LBTE solution against the conventional RTA. We find that the RTA strongly underestimates the thermal conductivity at low temperatures, where the off-diagonal structure of the collision matrix gives rise to relaxation eigenmodes that cannot be captured by independent single-mode lifetimes. We further find a pronounced anisotropy between the in-plane and out-of-plane thermal conductivities. A mode-resolved analysis shows that this anisotropy is carried by a small number of collective relaxation modes, and that the larger out-of-plane conductivity mainly originates from the stronger out-of-plane group-velocity contribution, rather than from a large difference in relaxation lifetimes. These results establish a direct connection between microscopic four-magnon relaxation, collision-matrix eigenmodes, and macroscopic anisotropic thermal transport in an altermagnetic material.

\section{Methods}
\label{sec:methods}

In this section, we describe the LBTE used to compute intrinsic magnon relaxation and thermal transport in $\alpha$-MnTe. The magnon spectrum is obtained from linear spin-wave theory, and the Holstein-Primakoff (HP) terms are used to derive the four-magnon transition rates through Fermi's golden rule\cite{holstein1940field}. These rates enter the Boltzmann collision integral; after linearization around thermal equilibrium and symmetrization of the deviation function, they define the symmetrized collision matrix of the LBTE. Diagonalizing this matrix yields the relaxation eigenmodes and relaxation rates, which are used to evaluate the thermal conductivity tensor and to benchmark the full collision-matrix solution against the RTA. The main equations are summarized below, while detailed spin-wave derivations and scattering vertices are provided in Apps.~\ref{appA} and \ref{appB}.

\subsection{Spin Hamiltonian and magnon spectrum}
\label{subsec:spin_hamiltonian}

We start from a spin Hamiltonian for the two-sublattice magnetic structure of
$\alpha$-MnTe\cite{szuszkiewicz2006Spin-wave}. The Hamiltonian is obtained as
\begin{equation}
H_0 =
\sum_{n=1}^{4}\sum_{\langle i,j\rangle_n}
J_n \mathbf{S}_i\cdot\mathbf{S}_j
-
K_{\rm an}\sum_i (S_i^z)^2 ,
\label{eq:H0}
\end{equation}
where $\langle i,j\rangle_n$ denotes the $n$th-neighbor exchange path in the NiAs-type lattice. We follow the exchange-path convention and parameter set of Refs.~\onlinecite{mu2019phonons,szuszkiewicz2006Spin-wave}: $J_1$ and $J_3$ connect Mn moments in adjacent layers and are antiferromagnetic, whereas $J_2$ and $J_4$ describe ferromagnetic exchange paths within the basal plane and between further-neighbor layers respectively. The exchange parameters are $J_1=21.5\,{\rm K}$, $J_2=-0.67\,{\rm K}$, $J_3=2.87\,{\rm K}$, and $J_4=-1.0\,{\rm K}$, with spin length $S=5/2$\cite{deltenre2021lattice}. Here $K_{\rm an}=0.26\,{\rm K}$  denotes the single-ion anisotropy\cite{szuszkiewicz2006Spin-wave}. For notational simplicity, we set $k_B=\hbar=1$ in the analytical expressions presented in this work, while the numerical results are reported after restoring physical units.

Applying the HP transformation and retaining the quadratic terms gives a bosonic Bogoliubov Hamiltonian. After Fourier transformation and Bogoliubov diagonalization, the noninteracting magnon Hamiltonian takes the form
\begin{equation}
H_0
=
\sum_{\mathbf{k}}
\epsilon_{\mathbf{k}}
\left(
\alpha_{\mathbf{k}}^\dagger\alpha_{\mathbf{k}}
+
\beta_{\mathbf{k}}^\dagger\beta_{\mathbf{k}}
\right),
\label{eq:H0_diag}
\end{equation}
where the two Bogoliubov branches are degenerate and
\begin{equation}
\epsilon_{\mathbf{k}}
=\sqrt{A_{\mathbf{k}}^2-B_{\mathbf{k}}^2}.
\label{eq:magnon_dispersion}
\end{equation}
The explicit expressions for $A_{\mathbf{k}}$ and $B_{\mathbf{k}}$ are given in App.~\ref{appA}.

\subsection{Four-magnon interactions and collision matrix}
\label{subsec:four_magnon}

The intrinsic magnon relaxation considered here arises from the quartic terms after the HP expansion. In the magnon eigenbasis, the quartic HP Hamiltonian contains both magnon-number-conserving two-in--two-out terms and number-nonconserving channels. For the $\alpha$-MnTe spin-wave dispersion considered here, the number-nonconserving channels are kinematically forbidden in the Fermi golden-rule collision integral, since they cannot simultaneously satisfy crystal-momentum conservation and energy conservation\cite{Harris1971dynamics}. We therefore retain only the number-conserving two-in--two-out processes.

The effective four-magnon interaction Hamiltonian can then be written in the compact
form
\begin{eqnarray}
H_{\rm scat}
&=&
\frac{1}{N}
\sum_{\nu_1\nu_2\nu_3\nu_4}
\sum_{\mathbf{k},\mathbf{p},\mathbf{r},\mathbf{s}}
\mathcal{M}^{\nu_1\nu_2;\nu_3\nu_4}_{\mathbf{k}\mathbf{p};\mathbf{r}\mathbf{s}}\,
\eta_{\nu_1\mathbf{k}}^\dagger
\eta_{\nu_2\mathbf{p}}^\dagger
\eta_{\nu_3\mathbf{r}}
\eta_{\nu_4\mathbf{s}}\,\nonumber\\
&\times&\delta_{\mathbf{k}+\mathbf{p},\,\mathbf{r}+\mathbf{s}+\mathbf{G}},
\label{eq:Hscat}
\end{eqnarray}
where $\eta_{\nu\mathbf{k}}$ denotes a Bogoliubov magnon operator in branch
$\nu=\alpha,\beta$, and $\mathcal{M}^{\nu_1\nu_2;\nu_3\nu_4}_{\mathbf{k}\mathbf{p};\mathbf{r}\mathbf{s}}$ is the branch-resolved four-magnon scattering matrix element, whose coefficients are given in App.~\ref{appB}. The reciprocal lattice vector $\mathbf{G}$ distinguishes Normal and Umklapp processes: $\mathbf{G}=0$ corresponds to Normal scattering, while $\mathbf{G}\neq0$ corresponds to Umklapp scattering\cite{peierls1929kinetischen,ziman1960electrons,akhiezer1968spin}.

The transition rate for $(\nu_1\mathbf{k},\nu_2\mathbf{p})\rightarrow(\nu_3\mathbf{r},\nu_4\mathbf{s})$ is obtained from Fermi's golden rule\cite{leon1954Quantum,leon1957The},
\begin{equation}
\begin{aligned}
W^{\nu_1\nu_2;\nu_3\nu_4}_{\mathbf{k}\mathbf{p};\mathbf{r}\mathbf{s}}
=&
\frac{2\pi}{ N^2}
\left|
\mathcal{M}^{\nu_1\nu_2;\nu_3\nu_4}_{\mathbf{k}\mathbf{p};\mathbf{r}\mathbf{s}}
\right|^2
\delta_{\mathbf{k}+\mathbf{p},\,\mathbf{r}+\mathbf{s}+\mathbf{G}}
\\
&\times
\delta(
\epsilon_{\mathbf{k}}+\epsilon_{\mathbf{p}}
-\epsilon_{\mathbf{r}}-\epsilon_{\mathbf{s}}
).
\end{aligned}
\label{eq:transition_rate}
\end{equation}
In the analysis below, the $\mathbf{G}=0$ and $\mathbf{G}\neq0$ contributions
are kept separately when comparing Normal-only and Normal-plus-Umklapp
collision operators.

\subsection{Linearized Boltzmann transport equation}
\label{subsec:lbte}

The two Bogoliubov magnon branches in Eq.~(\ref{eq:H0_diag}) are degenerate, and hence have the same equilibrium occupation. We therefore write $f^0_{\alpha\mathbf{k}}=f^0_{\beta\mathbf{k}}\equiv f^0_{\mathbf{k}}$ and suppress the branch index in the LBTE below.

In the absence of external forces, the semiclassical Boltzmann equation for
magnon wave packets is
\begin{equation}
\frac{\partial f_{\mathbf{k}}}{\partial t}
+
\mathbf{v}_{\mathbf{k}}\cdot\nabla_{\mathbf{r}} f_{\mathbf{k}}
=
\left.\frac{\partial f_{\mathbf{k}}}{\partial t}\right|_{\rm coll},
\label{eq:boltzmann_equation}
\end{equation}
where
\begin{equation}
\mathbf{v}_{\mathbf{k}}
=\nabla_{\mathbf{k}}\epsilon_{\mathbf{k}}
\end{equation}
is the magnon group velocity.

For the number-conserving four-magnon processes retained in this work, the
collision integral is
\begin{equation}
\begin{aligned}
\left.\frac{\partial f_{\mathbf{k}}}{\partial t}\right|_{\rm coll}
=&
\sum_{\mathbf{p},\mathbf{r},\mathbf{s}}
W_{\mathbf{k}\mathbf{p};\mathbf{r}\mathbf{s}}
\Big[
(1+f_{\mathbf{k}})(1+f_{\mathbf{p}})f_{\mathbf{r}}f_{\mathbf{s}}
\\
&\qquad
-
f_{\mathbf{k}}f_{\mathbf{p}}
(1+f_{\mathbf{r}})(1+f_{\mathbf{s}})
\Big].
\end{aligned}
\label{eq:collision_integral}
\end{equation}
Here the branch labels are suppressed for compactness, and
$W_{\mathbf{k}\mathbf{p};\mathbf{r}\mathbf{s}}$ denotes the transition rate
defined in Eq.~(\ref{eq:transition_rate}).

We first define the relaxation spectrum from the linearized
collision dynamics\cite{hua2024dynamics,chaput2013direct,cepellotti2016thermal,ward2009ab,broido2005lattice}. By writing
\begin{equation}
f_{\mathbf{k}}
=
f^0_{\mathbf{k}}
+
\delta f_{\mathbf{k}},
\qquad
f^0_{\mathbf{k}}
=
\frac{1}{\exp[\epsilon_{\mathbf{k}}/(k_BT)]-1},
\label{eq:bose_distribution}
\end{equation}
and introducing the rescaled deviation
\begin{equation}
\delta f_{\mathbf{k}}
=
\sqrt{f^0_{\mathbf{k}}(1+f^0_{\mathbf{k}})}
\,\phi_{\mathbf{k}},
\label{eq:rescaled_deviation}
\end{equation}
the collision integral is linearized and can be written as
\begin{equation}
\left.\frac{\partial f_{\mathbf{k}}}{\partial t}\right|_{\rm coll}^{(1)}
=
-
\sqrt{f^0_{\mathbf{k}}(1+f^0_{\mathbf{k}})}
\sum_{\mathbf{k}'}
\Omega_{\mathbf{k}\mathbf{k}'}
\phi_{\mathbf{k}'} .
\label{eq:linearized_collision}
\end{equation}
The rescaling in Eq.~(\ref{eq:rescaled_deviation}) brings the linearized
collision operator into a symmetric matrix $\Omega_{\mathbf{k}\mathbf{k}'}$.
For spatially uniform perturbations, Eq.~(\ref{eq:boltzmann_equation}) then
gives
\begin{equation}
\frac{\partial \phi_{\mathbf{k}}}{\partial t}
=
-
\sum_{\mathbf{k}'}
\Omega_{\mathbf{k}\mathbf{k}'}
\phi_{\mathbf{k}'} .
\label{eq:relaxation_dynamics}
\end{equation}
The relaxation eigenmodes are therefore defined by
\begin{equation}
\sum_{\mathbf{k}'}
\Omega_{\mathbf{k}\mathbf{k}'}
\psi_{i,\mathbf{k}'}
=
\lambda_i\psi_{i,\mathbf{k}},
\label{eq:collision_eigen}
\end{equation}
where $\lambda_i$ is the relaxation rate associated with the $i$th
collective mode, and $\psi_{i,\mathbf{k}}$ denotes its component in
momentum space. This definition of the relaxation spectrum is independent of the external temperature-gradient driving field.

We next use the same collision matrix to compute the linear response to a weak
temperature gradient. In the steady state, $\partial f_{\mathbf{k}}/\partial t=0$,
and the drift term is evaluated from the local equilibrium distribution:
\begin{equation}
\mathbf{v}_{\mathbf{k}}\cdot\nabla_{\mathbf{r}} f^0_{\mathbf{k}}
=
v_{\mathbf{k},a}
\frac{\partial f^0_{\mathbf{k}}}{\partial T}
\nabla_a T
=
\frac{\epsilon_{\mathbf{k}}v_{\mathbf{k},a}}{k_B T^2}
f^0_{\mathbf{k}}(1+f^0_{\mathbf{k}})
\nabla_a T ,
\label{eq:drift_term}
\end{equation}
where $a=x,y,z$ labels the direction of the applied temperature gradient.

Because the response is linear in $\nabla_aT$, we write the corresponding
rescaled deviation as
\begin{equation}
\phi_{\mathbf{k}}^{(a)}
=
-
\frac{\nabla_aT}{k_BT^2}
\varphi_{\mathbf{k}}^{(a)} .
\label{eq:response_decomposition}
\end{equation}
Substituting Eqs.~\eqref{eq:linearized_collision},
\eqref{eq:drift_term}, and \eqref{eq:response_decomposition} into the steady-state Boltzmann equation gives
\begin{equation}
X_{\mathbf{k},a}
=
\sum_{\mathbf{k}'}
\Omega_{\mathbf{k}\mathbf{k}'}
\varphi_{\mathbf{k}'}^{(a)} ,
\label{eq:sym_lbte}
\end{equation}
where
\begin{equation}
X_{\mathbf{k},a}
=
\epsilon_{\mathbf{k}}v_{\mathbf{k},a}
\sqrt{f^0_{\mathbf{k}}(1+f^0_{\mathbf{k}})}
\label{eq:heat_current_vector}
\end{equation}
is the heat-current driving vector. Thus the same symmetrized collision matrix
$\Omega$ determines both the relaxation spectrum in
Eq.~(\ref{eq:collision_eigen}) and the thermal response in
Eq.~(\ref{eq:sym_lbte}).

\subsection{Evaluation of thermal conductivity}
\label{subsec:kappa}

With the heat-current driving vector $X_{\mathbf{k},a}$, the thermal conductivity tensor can be obtained. For a gradient applied along direction $a=x,y,z$,
Eq.~(\ref{eq:sym_lbte}) gives the response function
\begin{equation}
\varphi^{(a)}
=
\Omega^{+}X_a ,
\label{eq:response_solution}
\end{equation}
where $\Omega^{+}$ denotes the pseudoinverse of the symmetrized collision
matrix after projecting out its zero modes.

The heat current along direction $a$ is
\begin{equation}
j_a
=
\frac{1}{V}
\sum_{\mathbf{k}}
\epsilon_{\mathbf{k}}v_{\mathbf{k},a}
\,\delta f_{\mathbf{k}}^{(a)} ,
\label{eq:heat_current}
\end{equation}
where, using Eq.~(\ref{eq:response_decomposition}),
\begin{equation}
\delta f_{\mathbf{k}}^{(a)}
=
\sqrt{f^0_{\mathbf{k}}(1+f^0_{\mathbf{k}})}
\,\phi_{\mathbf{k}}^{(a)}
=
-
\frac{\nabla_aT}{k_BT^2}
\sqrt{f^0_{\mathbf{k}}(1+f^0_{\mathbf{k}})}
\,\varphi_{\mathbf{k}}^{(a)} .
\label{eq:delta_f_response}
\end{equation}
Combining this expression with the definition of
$X_{\mathbf{k},a}$ in Eq.~(\ref{eq:heat_current_vector}), we obtain
\begin{equation}
j_a
=
-
\frac{\nabla_aT}{V k_B T^2}
\langle X_a|\varphi^{(a)}\rangle, 
\label{eq:current_response}
\end{equation}
with the inner product denoting summation over momentum states. Further using the linear-response relation $j_a=-\kappa_{a}\nabla_aT$, it gives
\begin{equation}
\kappa_{a}^{\rm LBTE}
=
\frac{1}{V k_B T^2}
\langle X_a|\Omega^{+}|X_a\rangle .
\label{eq:kappa_pseudoinverse}
\end{equation}

Equivalently, expanding the pseudoinverse in the eigenbasis of the
symmetrized collision matrix yields\cite{chaput2013direct}
\begin{equation}
\kappa_{a}^{\rm LBTE}
=
\frac{1}{V k_B T^2}
\sum_{\lambda_i>\lambda_{\rm cut}}
\frac{
\langle X_a|\psi_i\rangle
\langle \psi_i|X_a\rangle
}{\lambda_i}.
\label{eq:kappa_lbte}
\end{equation}
Here $\lambda_{\rm cut}$ is used to exclude the zero modes of the collision
matrix when constructing the pseudoinverse. Equivalently, the inverse
collision operator is applied only within the finite-rate subspace. For the zero modes that remain in the full collision operator, the projections onto the heat-current driving vectors vanish, and therefore these modes do not
contribute to Eq.~\eqref{eq:kappa_lbte}.  Removing these modes thus does not affect the
finite thermal conductivity.

For comparison, we also evaluate the RTA. In the RTA, the off-diagonal elements of the linearized collision matrix are neglected, so that each momentum mode relaxes independently. In the present
notation, this gives
\begin{equation}
\kappa_{a}^{\rm RTA}
=
\frac{1}{V k_B T^2}
\sum_{\mathbf{k}}
\epsilon_{\mathbf{k}}^2
v^2_{\mathbf{k},a}
f^0_{\mathbf{k}}(1+f^0_{\mathbf{k}})
\tau_{\mathbf{k}},
\label{eq:kappa_rta}
\end{equation}
where the single-mode relaxation time $\tau_{\mathbf{k}}$ is obtained from the
diagonal part of the linearized collision operator. The comparison between
Eqs.~(\ref{eq:kappa_lbte}) and (\ref{eq:kappa_rta}) quantifies the role of
off-diagonal collision processes and collective relaxation channels that are
absent in the single-mode approximation.

\section{Results and discussion}
\label{sec:results}

We now present the relaxation spectrum and thermal transport obtained from the symmetrized four-magnon collision matrix of $\alpha$-MnTe. We first examine how Normal and Umklapp scattering shape the low-lying relaxation spectrum. We then use the same collision matrix to compute the temperature-dependent thermal conductivity tensor and compare the full LBTE result with the RTA. Finally, we resolve the mode contributions to the thermal conductivity in order to identify the microscopic origin of the transport anisotropy.

\subsection{Relaxation spectrum}

\begin{figure}
    \centering
    \includegraphics[width=0.5\textwidth]{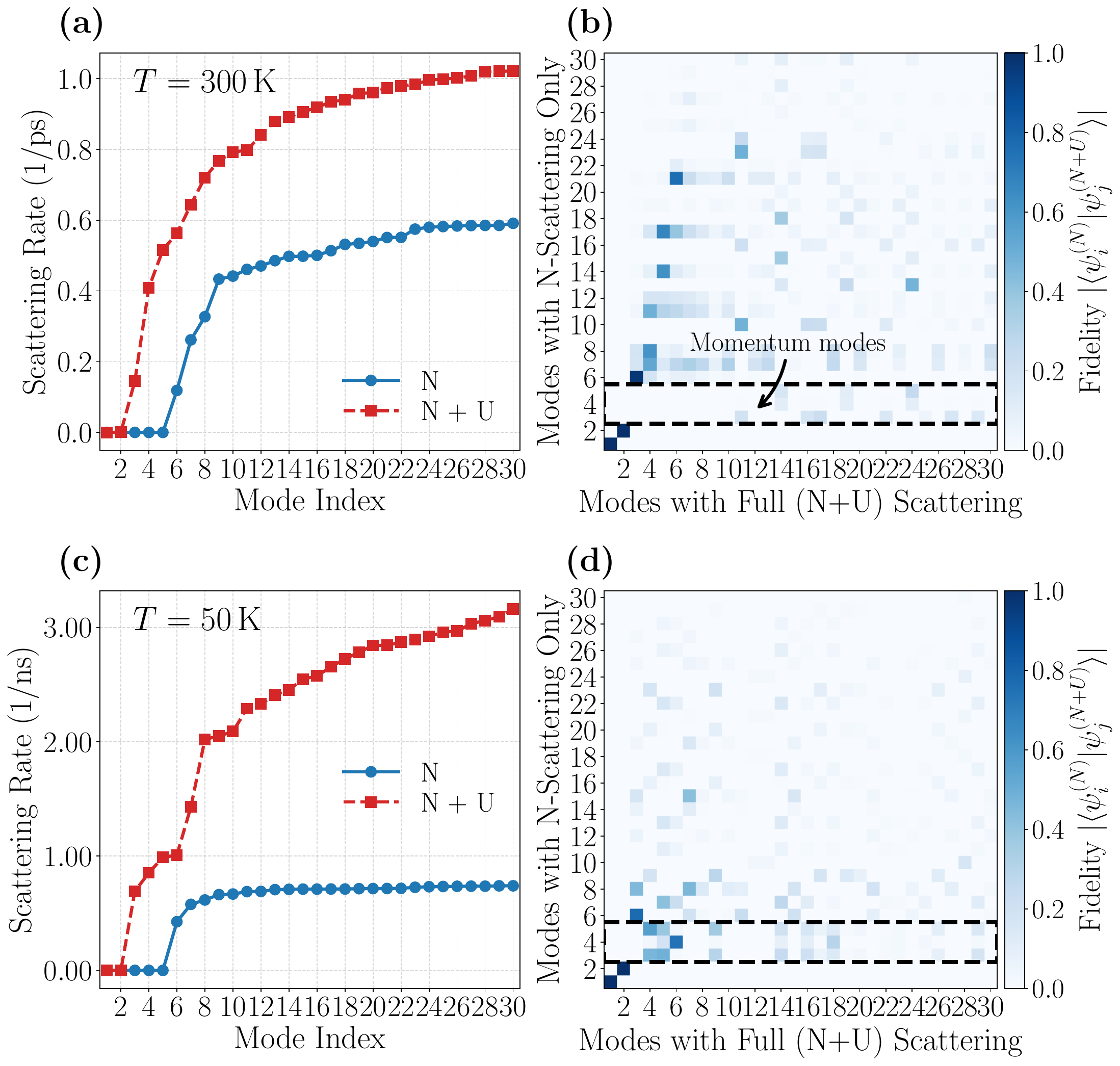}
    \caption{{\bf Low-lying relaxation spectrum and eigenmode correspondence.}
    (a) Scattering rates of the lowest 30 eigenmodes for the Normal-only (blue curves)
    and full Normal-plus-Umklapp (red curves) collision operators. The Normal-only operator
    contains five zero modes, whereas the full operator contains two zero
    modes. Including Umklapp scattering lifts the momentum-related zero modes
    and substantially enhances the finite relaxation rates throughout the
    displayed low-lying sector.
    (b) Absolute overlap matrix
    $|\langle \psi_i^{(N)}|\psi_j^{(N+U)}\rangle|$ at $300\,{\rm K}$.
    The absence of a simple shifted diagonal shows that Umklapp scattering
    does not merely relabel the Normal-only modes, but also mixes and reshapes the low-lying relaxation eigenmodes. 
    (c) and (d) Corresponding results at $T=50\,{\rm K}$. The overall relaxation
    rates are reduced to the $1\,{\rm ns}^{-1}$ scale, reflecting the
    strong suppression of four-magnon scattering at low temperature. However,
    the full Normal-plus-Umklapp rates remain visibly larger than the
    Normal-only rates for the displayed finite-rate modes, indicating that
    Umklapp scattering still modifies the slow relaxation sector. The overlap
    matrix in (d) further shows substantial mode mixing among the low-lying
    modes. The region enclosed by the two black dashed lines in panels (b) and (d) marks three momentum-related zero modes for the Normal-only collision operators.}
    \label{fig:1}
\end{figure}

To examine the microscopic relaxation structure of $\alpha$-MnTe, we first analyze the low-lying eigenvalue spectrum of the symmetrized collision matrix. Fig.~\ref{fig:1} compares the lowest 30 relaxation eigenmodes obtained from the Normal-only collision operator and from the full Normal-plus-Umklapp collision operator at two representative temperatures.

For the Normal-only spectrum (blue curves), it contains five zero
modes as shown in Figs.~\ref{fig:1}(a) and (c). Within the number-conserving four-magnon collision operator used here, these modes originate from the conservation of energy, magnon quasiparticle number, and the three components of crystal momentum \cite{daniel2019lecture}, as shown in App.~\ref{appC}. When Umklapp processes are included, crystal momentum is no longer conserved by the collision operator because momentum can be transferred to the lattice through a reciprocal lattice vector. Consequently, only the two zero modes associated with energy and magnon quasiparticle-number conservation remain, as shown by the red curves, while the three momentum-related zero modes acquire finite relaxation rates. 

Beyond this reduction of the zero modes, Fig.~\ref{fig:1}(a) shows that
Umklapp scattering strongly modifies the finite-rate part of the low-lying
spectrum at $300\,{\rm K}$. The full Normal-plus-Umklapp rates are
substantially larger than the corresponding Normal-only rates over most of the displayed finite-rate sector. For example, after the zero modes, the red curve rises rapidly to values of order $0.5$--$1\,{\rm ps}^{-1}$, whereas the Normal-only curve remains significantly lower over the same range of mode indices. This indicates that, at room temperature, thermally populated large-momentum states provide sufficient Umklapp phase space for momentum-relaxing processes to contribute directly to the low-lying relaxation spectrum.

The eigenmode correspondence in Fig.~\ref{fig:1}(b) confirms that the effect of Umklapp scattering is not limited to a simple shift of the mode index. If the full spectrum were obtained only by lifting the three momentum zero modes (indicated by the regions enclosed by the black dashed lines) without reshaping the finite-rate modes, one would expect a clear shifted high-fidelity diagonal in the overlap matrix $|\langle \psi_i^{(N)}|\psi_j^{(N+U)}\rangle|$ shown in this figure. Instead, it shows a more distributed overlap pattern. This indicates that the full Normal-plus-Umklapp eigenmodes are formed by mixing several Normal-only relaxation modes, rather than by a one-to-one relabeling of the Normal-only finite-rate spectrum.

We next compare this room-temperature behavior with the low-temperature
spectrum in Figs.~\ref{fig:1}(c) and (d). At $T=50\,{\rm K}$, the overall scale of the relaxation rates is strongly reduced: the displayed rates are of order $1/{\rm ns}$, about three orders of magnitude smaller than those in Fig.~\ref{fig:1}(a). This reduction reflects the suppressed thermal population and the reduced four-magnon scattering phase space at low temperature. At the level of the relaxation rates, Fig.~\ref{fig:1}(c) shows a qualitative similarity to the room-temperature case: the full Normal-plus-Umklapp rates remain clearly above the Normal-only rates for most of the displayed finite-rate modes. Thus, even at $50\,{\rm K}$, Umklapp scattering gives a visible correction to the slow relaxation sector.

However, the overlap matrix reveals that the microscopic origin of this
correction is different from that at $300\,{\rm K}$. In the room-temperature case, Fig.~\ref{fig:1}(b) shows that the momentum zero-mode character of the Normal-only operator is broadly redistributed among the full Normal-plus-Umklapp eigenmodes, without a single lowest finite-rate mode being dominated by this sector. By contrast, Fig.~\ref{fig:1}(d) shows that, at $50\,{\rm K}$, the three momentum-related zero modes of the Normal-only
operator, corresponding to rows $i=3$ to $5$ between the two black dashed horizontal lines, retain prominent projections onto the full Normal-plus-Umklapp modes in columns $j=3$, $5$, and $6$, although part of their weight is still distributed over higher modes. This indicates that Umklapp scattering does not completely remove the momentum-zero-mode character from the slow sector at low temperature. Instead, a sizable component of the Umklapp-lifted momentum modes remains in the low-rate window and hybridizes with the leading Normal relaxation modes.

Taken together, Fig.~\ref{fig:1} shows that Umklapp scattering modifies the low-lying relaxation spectrum at both temperatures, but in different ways. At $300\,{\rm K}$, the enlarged thermal phase space makes Umklapp scattering a strong relaxation channel: it broadly enhances the finite relaxation rates and redistributes the momentum zero modes  over several low-lying full eigenmodes. At $50\,{\rm K}$, the overall relaxation rates are much smaller, but Umklapp scattering still substantially increases the finite rates. More importantly, the momentum zero modes are not broadly dispersed; instead, a sizable part of them remain concentrated near the lowest finite-rate full modes.

\begin{figure}
    \centering
    \includegraphics[width=0.5\textwidth]{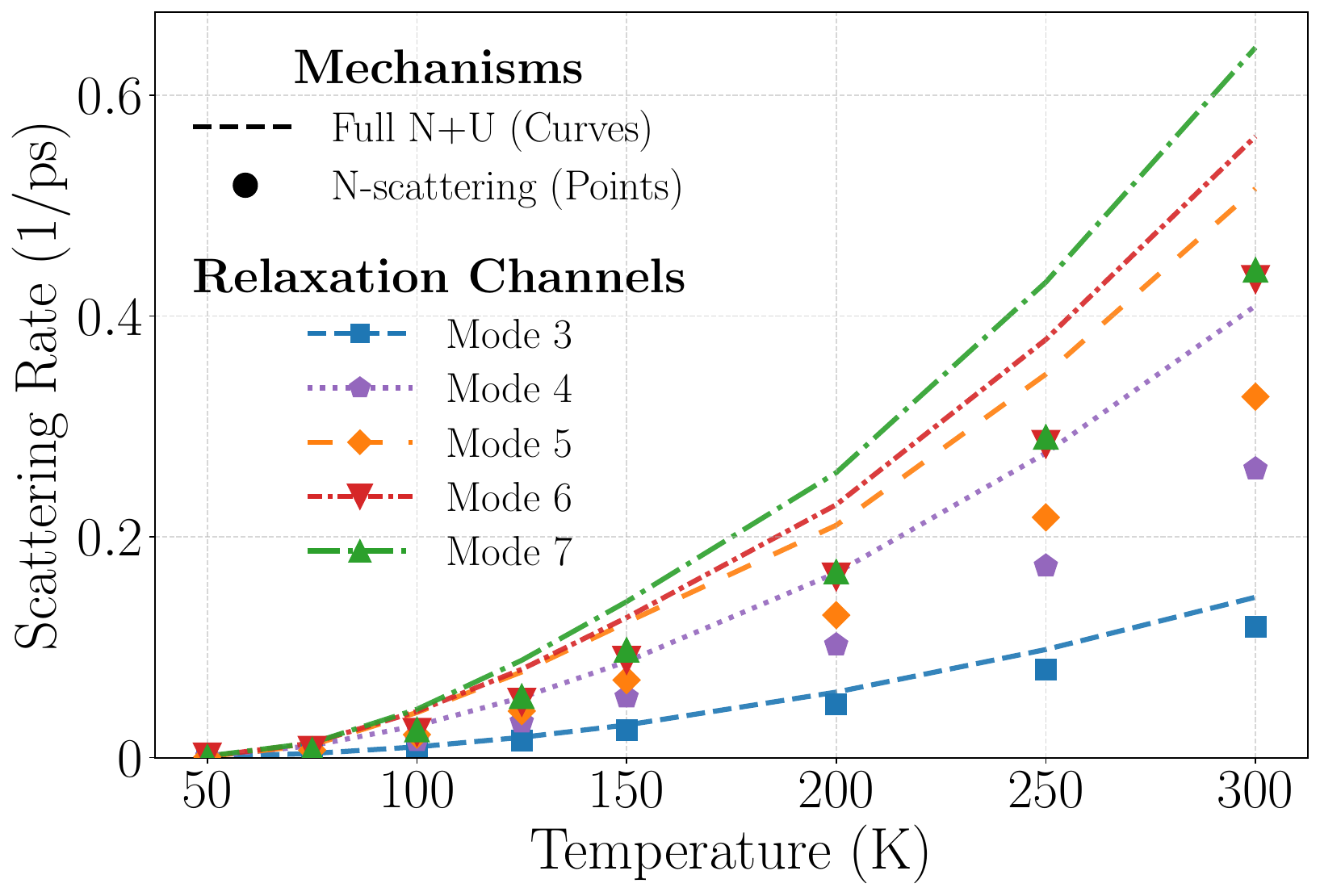}
    \caption{{\bf Temperature evolution of selected low-lying relaxation rates.} Scattering rates of Modes 3 to 7 as functions of temperature. The mode labels refer to the eigenmode indices of the full Normal-plus-Umklapp collision operator. Curves denote the full Normal-plus-Umklapp results. Points denote the corresponding Normal-only rates with the Normal-only mode index shifted by three, i.e., full mode $j$ is compared with Normal-only mode $j+3$, because the Normal-only operator contains three additional momentum-conservation zero modes. The systematic separation between curves and points shows that Umklapp scattering provides an additional relaxation channel for these modes, with a larger quantitative effect at elevated temperatures.}
    \label{fig:2}
\end{figure}

To further connect the two representative spectra in Fig.~\ref{fig:1}, we track the temperature evolution of selected low-lying relaxation modes. Fig.~\ref{fig:2} shows the scattering rates of Modes 3 to 7 from $50$ to $300\,{\rm K}$. Here the mode indices are defined with respect to the full Normal-plus-Umklapp collision operator. For the Normal-only data, the corresponding modes are compared after shifting the index by three, because the Normal-only operator contains three additional crystal-momentum zero modes. In other words, the point associated with full mode $j$ represents the Normal-only finite-rate mode with index $j+3$.

All displayed modes show an overall increase of the relaxation rate with temperature, reflecting the growth of the thermally available four-magnon scattering phase space. The comparison between the full Normal-plus-Umklapp curves and the shifted Normal-only points further shows the effect of Umklapp scattering: the full curves lie above the corresponding Normal-only points over the displayed temperature range, indicating that Umklapp scattering provides an additional relaxation channel. At low temperature, the absolute rates remain small, but the difference between the full and Normal-only results still exists. This is consistent with Figs.~\ref{fig:1}(c) and (d), where including Umklapp scattering substantially enhances the scattering rates in the low-rate sector. As temperature increases, the separation between the curves and points becomes more pronounced, confirming that Umklapp scattering becomes a quantitatively stronger relaxation channel at elevated temperatures. The enhancement is mode dependent, with different low-lying modes acquiring different Umklapp-induced corrections rather than a uniform shift of the relaxation spectrum.

\subsection{Thermal conductivity}

\begin{figure}
\centering
\includegraphics[width=0.5\textwidth]{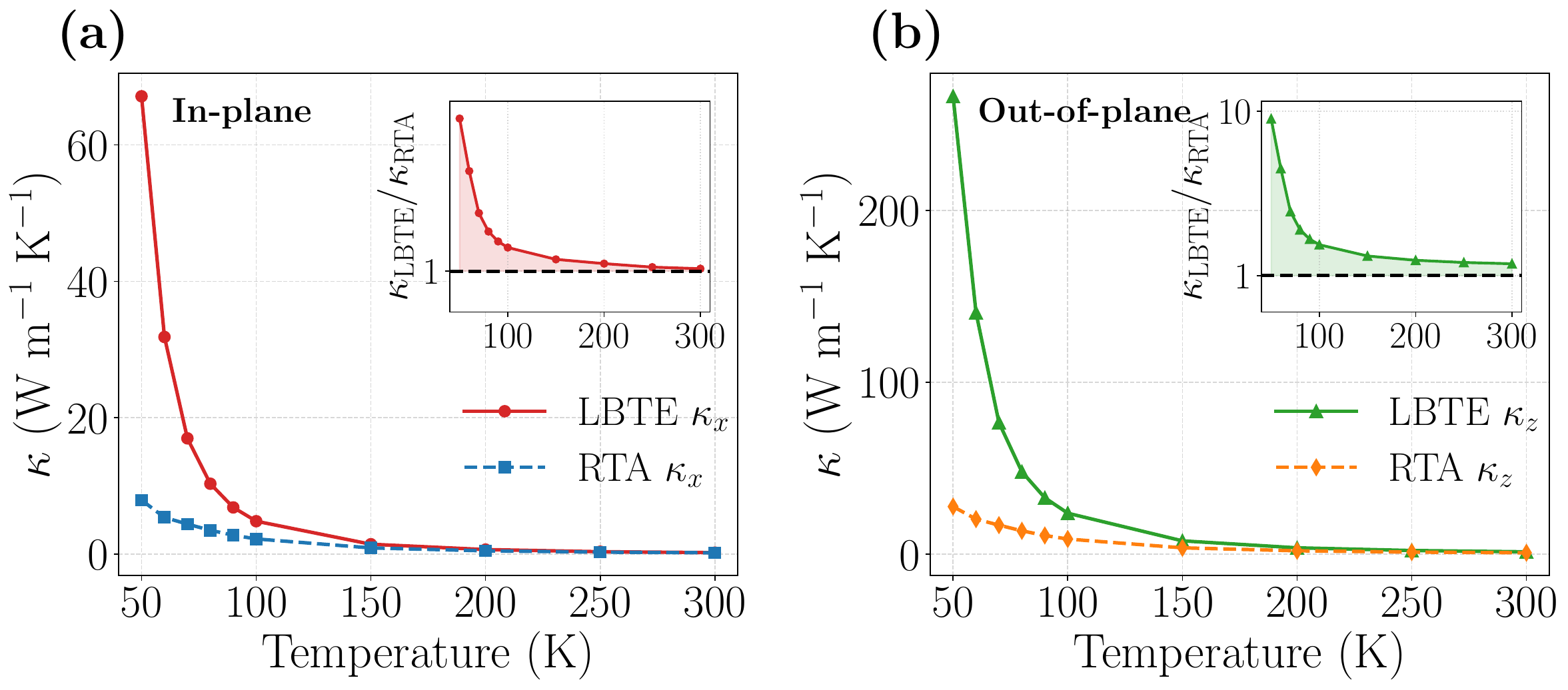}
\caption{{\bf Temperature dependence of the magnon thermal conductivity.}
(a) In-plane thermal conductivity $\kappa_{x}$ as a function of temperature.
The red solid curve denotes the full LBTE result, while the blue dashed curve denotes the RTA result. The inset shows the ratio $\kappa_{x}^{\rm LBTE}/\kappa_{x}^{\rm RTA}$, with the horizontal dashed line marking $\kappa_{x}^{\rm LBTE}/\kappa_{x}^{\rm RTA}=1$. The ratio remains above the unity line over the displayed temperature range and decreases toward unity as temperature increases.
(b) Out-of-plane thermal conductivity $\kappa_{z}$ as a function of temperature. The green solid curve denotes the full LBTE result, while the orange dashed curve denotes the RTA result. The inset shows the ratio $\kappa_{z}^{\rm LBTE}/\kappa_{z}^{\rm RTA}$, with the horizontal dashed line marking $\kappa_{z}^{\rm LBTE}/\kappa_{z}^{\rm RTA}=1$. The ratio also remains above unity over the displayed temperature range and decreases strongly with increasing temperature. In both directions, the full LBTE
conductivity exceeds the RTA result, with the largest enhancement occurring at low temperature. The out-of-plane conductivity remains larger than the in-plane conductivity over the whole plotted temperature range.}
\label{fig:3}
\end{figure}

We now connect the relaxation spectrum to the macroscopic magnon thermal conductivity. Fig.~\ref{fig:3} shows the temperature dependence of the
in-plane thermal conductivity $\kappa_{x}$ and the out-of-plane thermal conductivity $\kappa_{z}$. The solid curves denote the full LBTE results [see Eq.~\eqref{eq:kappa_pseudoinverse}], while the dashed curves denote the
RTA results [see Eq.~\eqref{eq:kappa_rta}]. The insets in Figs.~\ref{fig:3}(a) and (b) show the corresponding ratios $\kappa_{x}^{\rm LBTE}/\kappa_{x}^{\rm RTA}$ and $\kappa_{z}^{\rm LBTE}/\kappa_{z}^{\rm RTA}$, respectively.

In Fig.~\ref{fig:3}, both the in-plane and out-of-plane conductivities show the same qualitative temperature dependence. At low temperature, the full LBTE conductivities
are much larger than the RTA results in both directions. At $T=50\,{\rm K}$, $\kappa_{x}^{\rm LBTE}=67.1\,{\rm W\,m^{-1}K^{-1}}$, whereas $\kappa_{x}^{\rm RTA}=7.9\,{\rm W\,m^{-1}K^{-1}}$. For the out-of-plane response, $\kappa_{z}^{\rm LBTE}=266.4\,{\rm W\,m^{-1}K^{-1}}$, while $\kappa_{z}^{\rm RTA}=27.8\,{\rm W\,m^{-1}K^{-1}}$. Thus, in the low-temperature regime, the RTA strongly underestimates the magnon thermal conductance for both in-plane and out-of-plane transport.

As temperature increases, the full LBTE conductivities along both directions are rapidly suppressed, reflecting the enhanced four-magnon relaxation rates. Around $T=100\,{\rm K}$, the full LBTE values decrease to $\kappa_{x}^{\rm LBTE}=4.8\,{\rm W\,m^{-1}K^{-1}}$ and $\kappa_{z}^{\rm LBTE}=23.9\,{\rm W\,m^{-1}K^{-1}}$. At the same temperature, the corresponding RTA values are reduced to $\kappa_{x}^{\rm RTA}=2.3\,{\rm W\,m^{-1}K^{-1}}$ and $\kappa_{z}^{\rm RTA}=8.9\,{\rm W\,m^{-1}K^{-1}}$. More generally, the full temperature dependence in Fig.~\ref{fig:3} shows that the RTA curves remain below the full LBTE curves throughout the plotted temperature range. Consistently, the insets show that both $\kappa_{x}^{\rm LBTE}/\kappa_{x}^{\rm RTA}$ and $\kappa_{z}^{\rm LBTE}/\kappa_{z}^{\rm RTA}$ remain above unity and approach unity from above as temperature increases. At $300\,{\rm K}$, the ratios become 1.12 and 1.64, respectively. Therefore, the discrepancy between the full LBTE and RTA results becomes progressively smaller at high temperature, but the RTA continues to underestimate the thermal conductivity in both directions throughout the displayed range.

It is noted that although the temperature dependence is similar in the two directions, the magnitude of the conductivity is strongly anisotropic. Comparing
Figs.~\ref{fig:3}(a) and \ref{fig:3}(b), the out-of-plane conductivity $\kappa_z$ remains larger than the in-plane conductivity $\kappa_x$ over the whole temperature range. At $T=50\,{\rm K}$, the full LBTE values give $\kappa_z/\kappa_x=266.4/67.1\approx4.0$, and at $300\,{\rm K}$, the ratio increases to $\kappa_z/\kappa_x=1.46/0.24\approx6.1$. Since the full LBTE conductivity depends on both the relaxation eigenvalues of the collision matrix and the projection of the
heat-current driving vector onto the corresponding eigenmodes, the microscopic origin of this anisotropy requires a mode-resolved analysis, which we present next.

\subsection{Mode-resolved origin of the transport anisotropy}

\begin{figure}
    \centering
    \includegraphics[width=0.5\textwidth]{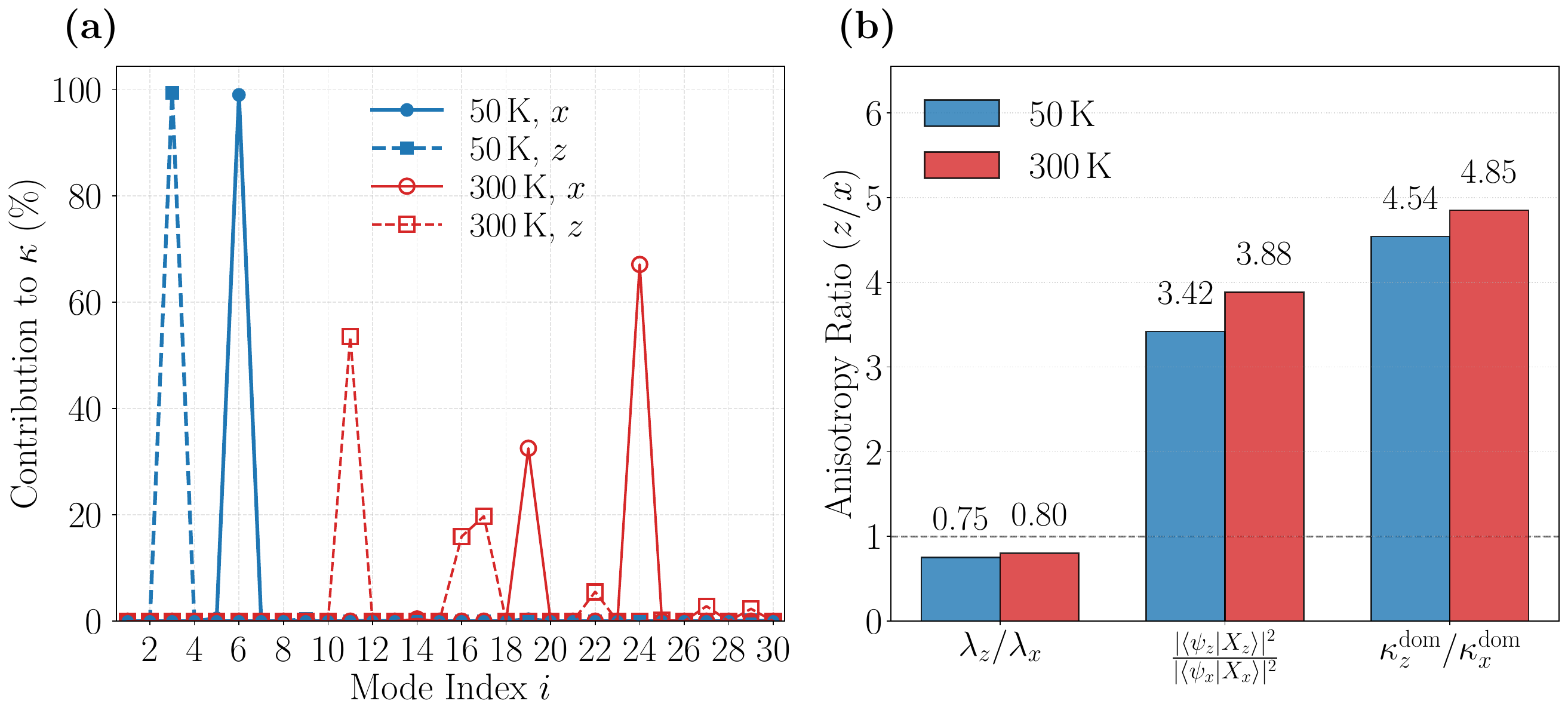}
  \caption{\textbf{Mode-resolved magnon thermal transport and anisotropy in
$\alpha$-MnTe.}
(a) Mode-resolved percentage contributions to the full LBTE thermal
conductivity at $50\,{\rm K}$ (blue) and $300\,{\rm K}$ (red). Solid lines
with circles denote the in-plane component $\kappa_{x}$, while dashed lines
with squares denote the out-of-plane component $\kappa_{z}$. The sharp peaks
show that the thermal conductivity is carried mainly by a small number of
relaxation eigenmodes, and that the dominant modes selected by the in-plane
and out-of-plane heat currents are different.
(b) Decomposition of the dominant-mode contribution to the transport
anisotropy. For the dominant $x$- and $z$-carrying modes, the bars show the
relaxation-eigenvalue ratio $\lambda_z/\lambda_x$, the heat-current projection
ratio $|\langle \psi_z|X_z\rangle|^2/|\langle \psi_x|X_x\rangle|^2$,
and the resulting dominant-mode conductivity ratio $\kappa_z^{\rm dom}/\kappa_x^{\rm dom}$. The horizontal dashed line marks the isotropic value of unity. Since
$\kappa_\alpha^{\rm dom}\propto|\langle \psi_\alpha|X_\alpha\rangle|^2/\lambda_\alpha$, the conductivity ratio is mainly controlled by the heat-current projection ratio rather than by the modest difference in relaxation eigenvalues.}
    \label{fig:4}
\end{figure}

To identify how the anisotropic thermal conductivity is distributed among the
relaxation eigenmodes, we decompose the full LBTE conductivity into
mode-resolved contributions. From Eq.~\eqref{eq:kappa_lbte}, the contribution
of the $i$th relaxation eigenmode to the diagonal component
$\kappa_{\alpha}$ is given by
\begin{equation}
\kappa_{\alpha}^{(i)}
=
\frac{1}{V k_B T^2}
\frac{
|\langle \psi_i|X_\alpha\rangle|^2
}{\lambda_i},
\label{eq:kappa_mode_resolved}
\end{equation}
where $X_\alpha$ is the heat-current driving vector [see Eq.~\eqref{eq:heat_current_vector}] for transport along direction $\alpha=x,z$, and $\lambda_i$ is the relaxation eigenvalue of the collision matrix. By evaluating $\kappa_{\alpha}^{(i)}$ for each eigenmode, we can determine which relaxation modes carry the dominant part of the total thermal conductivity.

Fig.~\ref{fig:4}(a) shows that the thermal conductivity is concentrated in a
small number of collision eigenmodes. At $50\,{\rm K}$, the in-plane
conductivity is dominated by the mode at $i=6$, as shown by the blue solid
peak, whereas the out-of-plane conductivity is dominated by a different mode
at $i=3$, as shown by the blue dashed peak. Comparing with the overlap
analysis in Fig.~\ref{fig:1}(d), both dominant low-temperature heat-carrying
modes retain visible overlap with the momentum-zero-mode sector of the
Normal-only collision operator. This is seen from their substantial weight in
the region between the two black horizontal dashed lines in Fig.~\ref{fig:1}(d), which marks the three momentum-related Normal-only zero modes. This behavior is consistent with the structure of the heat-current driving vector $X_{\mathbf{k},a}$, which has a momentum-like form in the low-energy regime and closely resembles the corresponding momentum-conservation zero mode, as discussed in App.~\ref{appD}. The
dominant transport modes at low temperature can therefore be understood as
Umklapp-lifted momentum modes that couple efficiently to the thermal driving field.

A related, although less pronounced, structure is also visible at $300\,{\rm K}$. The red solid curve shows that $\kappa_{x}$ is dominated by the modes at $i=19$ and $24$ within the displayed low-lying sector, while the red dashed curve shows that $\kappa_{z}$ receives its main contribution from lower-index modes, especially at $i=11$ with additional weight at $i=16$, $17$, and $22$. According to Fig.~\ref{fig:1}(b), the momentum-zero-mode character is more broadly redistributed at this temperature, but the dominant heat-carrying modes still retain visible momentum-like weight. Thus, even at $300\,{\rm K}$, the dominant heat-carrying modes remain tied to the momentum-zero-mode sector of the Normal-only collision operator.

Having identified the dominant heat-carrying modes, we next ask why the
out-of-plane conductivity is larger than the in-plane one. Fig.~\ref{fig:4}(b)
compares the dominant $z$-carrying mode with the dominant $x$-carrying mode
by separating the relaxation-rate contribution from the heat-current-driving
contribution. The first group of bars shows the eigenvalue ratio
$\lambda_z/\lambda_x$. This ratio is smaller than unity at both temperatures,
with $\lambda_z/\lambda_x\simeq0.75$ at $50\,{\rm K}$ and $\simeq0.8$ at
$300\,{\rm K}$. Therefore, the dominant $z$-carrying mode has a slightly
smaller relaxation rate than the dominant $x$-carrying mode. However, this effect is modest.
By contrast, the heat-current projection ratio
$|\langle\psi_z|X_z\rangle|^2/|\langle\psi_x|X_x\rangle|^2$, shown by the
second group of bars in Fig.~\ref{fig:4}(b), is much larger:
it is about $3.42$ at $50\,{\rm K}$ and increases to about $3.88$ at
$300\,{\rm K}$. These projection ratios are much larger than the modest variation in
$\lambda_z/\lambda_x$, indicating that the enhanced $\kappa_z$ is governed
primarily by the stronger heat-current projection. Since the directional
dependence of the driving vector
$X_{\mathbf{k},a}=\epsilon_{\mathbf{k}}v_{\mathbf{k},a}
\sqrt{f_{\mathbf{k}}^0(1+f_{\mathbf{k}}^0)}$
enters through the group-velocity component $v_{\mathbf{k},a}$, the larger
$\kappa_z$ mainly reflects the stronger out-of-plane group-velocity
contribution to the thermal driving vector, while the smaller relaxation rate of the
dominant $z$-carrying mode provides only a secondary correction.

\section{Summary}
\label{sec:summary}

In summary, we have presented a full collision-matrix solution of the LBTE for intrinsic magnon thermal transport in $\alpha$-MnTe. Starting from the four-magnon interaction, we constructed the symmetrized collision matrix and analyzed both its
relaxation eigenmodes and the resulting thermal conductivity. This approach goes beyond the RTA by retaining the off-diagonal collision processes that couple different momentum states into collective relaxation eigenmodes.

The relaxation spectrum reveals the distinct roles of Normal and Umklapp scattering. For the Normal-only collision operator, five zero modes appear due to energy, magnon-number, and crystal-momentum conservation. Including Umklapp processes lifts the three momentum-related zero modes and leaves only the two zero modes associated with energy and magnon-number conservation. At $300\,{\rm K}$, Umklapp scattering strongly enhances the finite relaxation rates and mixes the low-lying eigenmodes. At $50\,{\rm K}$, although the absolute relaxation rates are much smaller, the Umklapp scattering 
still visibly modifies the lowest relaxation modes.

Using the same collision matrix, we find that the full LBTE thermal conductivity is
strongly enhanced over the RTA at low temperature. At $50\,{\rm K}$, $\kappa_{x}^{\rm LBTE}=67.1\,{\rm W\,m^{-1}K^{-1}}$ and $\kappa_{z}^{\rm LBTE}=266.4\,{\rm W\,m^{-1}K^{-1}}$, more than an order of magnitude larger than the corresponding RTA values. With increasing temperature, the conductivities are rapidly suppressed, consistent with the growth of the four-magnon relaxation rates. Nevertheless, the RTA remains below the full LBTE result throughout the studied temperature range, although the
relative discrepancy becomes less pronounced at higher temperatures. A mode-resolved decomposition shows that the thermal conductivity is carried mainly by a small number of collective relaxation modes. These dominant modes remain tied to the momentum-zero-mode sector of the Normal-only collision operator, indicating that Umklapp-lifted momentum-like modes form the main heat-carrying channels. We also find a pronounced transport anisotropy, $\kappa_{z}>\kappa_{x}$, throughout the studied temperature range. Comparing the dominant $x$- and $z$-carrying modes further shows that this
anisotropy is primarily caused by the stronger out-of-plane group-velocity
contribution to the heat-current driving, while the difference in their relaxation lifetimes plays only a secondary role.

\begin{acknowledgments}
The authors thank Cheng-yun Hua for his valuable suggestions. This work is supported by the National Natural Science Foundation of China (Grants No. 12204399).
\end{acknowledgments}

\appendix

\section{Derivation of the free Hamiltonian}
\label{appA}
In this appendix, we collect the quadratic spin-wave Hamiltonian, magnon
dispersion, and Bogoliubov coefficients used in the main text.

The magnetic structure of $\alpha$-MnTe consists of two sublattices, denoted
by $a$ and $b$. Starting from Eq.~\eqref{eq:H0}, we use the HP transformation
\cite{holstein1940field}
\begin{align}
S_{a,i}^{+} &= \sqrt{2S-a_i^\dagger a_i}\,a_i,
&
S_{b,j}^{+} &= b_j^\dagger \sqrt{2S-b_j^\dagger b_j},
\\
S_{a,i}^{-} &= a_i^\dagger \sqrt{2S-a_i^\dagger a_i},
&
S_{b,j}^{-} &= \sqrt{2S-b_j^\dagger b_j}\,b_j,
\\
S_{a,i}^{z} &= S-a_i^\dagger a_i,
&
S_{b,j}^{z} &= b_j^\dagger b_j-S .
\end{align}

Keeping terms up to quadratic order and omitting an irrelevant constant
energy shift, we obtain the bosonic Bogoliubov Hamiltonian
\begin{equation}
H_0
=\frac{1}{2}
\sum_{\mathbf{k}}
\Psi_{\mathbf{k}}^\dagger
\mathcal{H}_{\mathbf{k}}
\Psi_{\mathbf{k}},
\label{eq:app_H0_matrix}
\end{equation}
where
\begin{equation}
\Psi_{\mathbf{k}}^\dagger
=
\begin{pmatrix}
a_{\mathbf{k}}^\dagger &
b_{\mathbf{k}}^\dagger &
a_{\mathbf{k}} &
b_{\mathbf{k}}
\end{pmatrix},
\end{equation}
and
\begin{equation}
\mathcal{H}_{\mathbf{k}}
=
\begin{pmatrix}
A_{\mathbf{k}} & 0 & 0 & B_{\mathbf{k}} \\
0 & A_{\mathbf{k}} & B_{\mathbf{k}} & 0 \\
0 & B_{\mathbf{k}} & A_{\mathbf{k}} & 0 \\
B_{\mathbf{k}} & 0 & 0 & A_{\mathbf{k}}
\end{pmatrix}.
\end{equation}

The matrix elements are
\begin{equation}
\begin{aligned}
A_{\mathbf{k}}
&=
2S\left[
z_1J_1+z_3J_3
+z_2J_2\left(1-\gamma_{\mathbf{k}}^2\right)
+z_4J_4\left(1-\gamma_{\mathbf{k}}^4\right)
-K
\right],
\\
B_{\mathbf{k}}
&=
2S\left[
z_1J_1\gamma_{\mathbf{k}}^1
+z_3J_3\gamma_{\mathbf{k}}^3
\right].
\end{aligned}
\label{eq:app_AB}
\end{equation}
Here $z_n$ is the coordination number for the $n$th-neighbor exchange path.
The structure factors for the NiAs-type lattice are
\begin{equation}
\begin{aligned}
\gamma_{\mathbf{k}}^1
&=
\cos\left(\frac{k_z c}{2}\right),
\\
\gamma_{\mathbf{k}}^2
&=
\frac{1}{3}
\left[
\cos(k_xa)
+
2\cos\left(\frac{k_xa}{2}\right)
\cos\left(\frac{\sqrt{3}k_ya}{2}\right)
\right],
\\
\gamma_{\mathbf{k}}^3
&=
\gamma_{\mathbf{k}}^1\gamma_{\mathbf{k}}^2,
\qquad
\gamma_{\mathbf{k}}^4
=
\cos(k_zc).
\end{aligned}
\label{eq:app_structure_factors}
\end{equation}
Here $a=4.15\,$\AA\, and $c=6.71\,$\AA\, are the lattice
constants. 

To diagonalize Eq.~\eqref{eq:app_H0_matrix}, we introduce the Bogoliubov
transformation
\begin{equation}
\begin{aligned}
a_{\mathbf{k}}
&=
u_{\mathbf{k}}\alpha_{\mathbf{k}}
+
v_{\mathbf{k}}\beta_{\mathbf{k}}^\dagger,
\\
b_{\mathbf{k}}^\dagger
&=
v_{\mathbf{k}}\alpha_{\mathbf{k}}
+
u_{\mathbf{k}}\beta_{\mathbf{k}}^\dagger .
\end{aligned}
\label{eq:app_bogoliubov}
\end{equation}
The diagonalization condition, together with the bosonic Bogoliubov normalization condition $u_{\mathbf{k}}^2-v_{\mathbf{k}}^2=1$, which preserves the canonical commutation relation of the transformed bosonic operators, gives
\begin{equation}
\begin{aligned}
u_{\mathbf{k}}
&=
-\frac{1}{\sqrt{2}}
\left[
1+
\frac{A_{\mathbf{k}}}
{\sqrt{A_{\mathbf{k}}^2-B_{\mathbf{k}}^2}}
\right]^{1/2},
\\
v_{\mathbf{k}}
&=
u_{\mathbf{k}}x_{\mathbf{k}},
\end{aligned}
\label{eq:app_uv}
\end{equation}
with
\begin{equation}
x_{\mathbf{k}}
=\frac{A_{\mathbf{k}}}{B_{\mathbf{k}}}
-\left[
\frac{A_{\mathbf{k}}^2}{B_{\mathbf{k}}^2}
-1
\right]^{1/2}.
\end{equation}
After dropping the additive constant generated by normal ordering, the
diagonalized Hamiltonian becomes
\begin{equation}
H_0
=\sum_{\mathbf{k}}
\epsilon_{\mathbf{k}}
\left(
\alpha_{\mathbf{k}}^\dagger\alpha_{\mathbf{k}}
+
\beta_{\mathbf{k}}^\dagger\beta_{\mathbf{k}}
\right),
\end{equation}
with the two degenerate magnon branches
\begin{equation}
\epsilon_{\mathbf{k}}
=\sqrt{A_{\mathbf{k}}^2-B_{\mathbf{k}}^2}.
\label{eq:app_dispersion}
\end{equation}

\section{Magnon-magnon interaction Hamiltonian\label{appB}}
In App.~\ref{appA}, the noninteracting magnon spectrum was obtained after substituting the leading HP representation into the spin Hamiltonian. To obtain the magnon-magnon interaction, we now retain the four-boson terms of the Hamiltonian. These terms arise from the next-to-leading contributions in the transverse spin operators, together with the longitudinal spin components.
Explicitly, we use
\begin{align}
    &S_{a,i}^{+}\approx\sqrt{2S}(1-\dfrac{a_i^{\dagger}a_i}{4S})a_{i},\quad S_{b,j}^{+}\approx b_j^{\dagger}\sqrt{2S}(1-\dfrac{b_j^{\dagger}b_j}{4S}),\\
    &S_{a,i}^{-}\approx a_i^{\dagger}\sqrt{2S}(1-\dfrac{a_i^{\dagger}a_i}{4S}),\quad S_{b,j}^{-}\approx\sqrt{2S}(1-\dfrac{b_j^{\dagger}b_j}{4S})b_{j}.
\end{align}

Then, substituting these expressions into the spin Hamiltonian and keeping the
terms with four bosonic operators gives the magnon-magnon interaction
Hamiltonian. For the spin-wave dispersion considered here, the magnon-number-nonconserving channels are kinematically forbidden in the Fermi's golden rule collision integral, since they cannot simultaneously satisfy energy and crystal-momentum conservation\cite{Harris1971dynamics}. We therefore retain only the number-conserving two-in-two-out processes. After Fourier transformation and Bogoliubov transformation, these interaction terms can be written as
\begin{align}
    H_{\xi}^{'}=&\dfrac{-J_{\xi}z_{\xi}}{2N}\sum_{\mathbf{k}\mathbf{p}\mathbf{s}\mathbf{r}}\delta(\mathbf{k}+\mathbf{p}-\mathbf{s}-\mathbf{r}+G)u_\mathbf{k}u_\mathbf{p}u_\mathbf{s}u_\mathbf{r}\nonumber\\
    &\qquad\big(\Phi_{\xi}\alpha_\mathbf{k}^{\dagger}\alpha_\mathbf{r}\beta_\mathbf{s}^{\dagger}\beta_\mathbf{p}+\Phi_{\xi}^{'}\alpha_\mathbf{k}^{\dagger}\alpha_\mathbf{p}^{\dagger}\alpha_\mathbf{s}\alpha_\mathbf{r}+\Phi_{\xi}^{''}\beta_\mathbf{k}\beta_\mathbf{p}\beta_\mathbf{s}^{\dagger}\beta_\mathbf{r}^{\dagger}\big),\\
    H_{\zeta}^{'}=&\dfrac{-J_{\zeta}z_{\zeta}}{4N}\sum_{\mathbf{k}\mathbf{p}\mathbf{s}\mathbf{r}}\delta(\mathbf{k}+\mathbf{p}-\mathbf{s}-\mathbf{r}+G)u_\mathbf{k}u_\mathbf{p}u_\mathbf{s}u_\mathbf{r}\nonumber\\
    &\big(\Phi_{\zeta}\alpha_\mathbf{k}^{\dagger}\alpha_\mathbf{r}\beta_\mathbf{s}^{\dagger}\beta_\mathbf{p}+\Phi_{\zeta}^{'}\alpha_\mathbf{k}^{\dagger}\alpha_\mathbf{p}^{\dagger}\alpha_\mathbf{s}\alpha_\mathbf{r}+\Phi_{\zeta}^{''}\beta_\mathbf{k}\beta_\mathbf{p}\beta_\mathbf{s}^{\dagger}\beta_\mathbf{r}^{\dagger}\big).
    \label{eq:HxiHzeta}
\end{align}
Here, $\xi=1,3$ labels the contributions from the $J_1$ and $J_3$
exchange interactions, while $\zeta=2,4$ labels those from the $J_2$ and
$J_4$ exchange interactions. As for the form factors
$\Phi_{\xi}$, $\Phi_{\xi}^{'}$, $\Phi_{\xi}^{''}$ and
$\Phi_{\zeta}$, $\Phi_{\zeta}^{'}$, $\Phi_{\zeta}^{''}$, they are obtained
after the Fourier and Bogoliubov transformations and are given by
\cite{Harris1971dynamics}:
\begin{align}
    \hspace{-2.5cm}
    \Phi_{\xi}=&\big(4\gamma_{\mathbf{s}-\mathbf{p}}^{\xi}+4\gamma_{\mathbf{r}-\mathbf{p}}^{\xi}x_\mathbf{s}x_\mathbf{r}+4\gamma_{\mathbf{s}-\mathbf{k}}^{\xi}x_\mathbf{k}x_\mathbf{p}+4\gamma_{\mathbf{r}-\mathbf{k}}^{\xi}x_\mathbf{k}x_\mathbf{p}x_\mathbf{s}x_\mathbf{r}\nonumber\\
    +&2\gamma_{\mathbf{s}-\mathbf{p}-\mathbf{k}}^{\xi}x_\mathbf{k}+2\gamma_{\mathbf{r}-\mathbf{p}-\mathbf{k}}^{\xi}x_\mathbf{k}x_\mathbf{s}x_\mathbf{r}+2\gamma_\mathbf{p}^{\xi}x_\mathbf{s}+2\gamma_{\mathbf{k}}^{\xi}x_\mathbf{k}x_\mathbf{p}x_\mathbf{s}\nonumber\\
    +&2\gamma_{\mathbf{s}+\mathbf{r}-\mathbf{p}}^{\xi}x_\mathbf{r}+2\gamma_{\mathbf{s}+\mathbf{r}-\mathbf{k}}^{\xi}x_\mathbf{k}x_\mathbf{p}x_\mathbf{r}+2\gamma_\mathbf{s}^{\xi}x_\mathbf{p}+2\gamma_\mathbf{r}^{\xi}x_\mathbf{p}x_\mathbf{s}x_\mathbf{r}\big),\\
    \Phi_{\xi}^{'}=&\Phi_{\xi}^{''}=\big(4\gamma_{\mathbf{s}-\mathbf{p}}^{\xi}x_\mathbf{p}x_\mathbf{s}+\gamma_{\mathbf{r}-\mathbf{k}-\mathbf{p}}^{\xi}x_\mathbf{k}x_\mathbf{p}x_\mathbf{r}+\gamma_\mathbf{p}^{\xi}x_\mathbf{p}\nonumber\\
    &\qquad+\gamma_{\mathbf{r}+\mathbf{s}-\mathbf{p}}^{\xi}x_\mathbf{p}x_\mathbf{s}x_\mathbf{r}+\gamma_\mathbf{r}^{\xi}x_\mathbf{r}\big),\\
    \Phi_{\zeta}=&\left(4\gamma_{\mathbf{s}-\mathbf{p}}^{\zeta}-\gamma_{\mathbf{s}-\mathbf{k}-\mathbf{p}}^{\zeta}-\gamma_{-\mathbf{p}}^{\zeta}-\gamma_{\mathbf{s}+\mathbf{r}-\mathbf{p}}^{\zeta}-\gamma_{\mathbf{s}}^{\zeta}\right)x_\mathbf{p}x_\mathbf{s}\nonumber\\
    &+\left(4\gamma_{\mathbf{r}-\mathbf{p}}^{\zeta}-\gamma_{\mathbf{r}-\mathbf{k}-\mathbf{p}}^{\zeta}-\gamma_{-\mathbf{p}}^{\zeta}-\gamma_{\mathbf{s}+\mathbf{r}-\mathbf{p}}^{\zeta}-\gamma_{\mathbf{r}}^{\zeta}\right)x_\mathbf{p}x_\mathbf{s}\nonumber\\
    &+\left(4\gamma_{\mathbf{s}-\mathbf{k}}^{\zeta}-\gamma_{\mathbf{s}-\mathbf{k}-\mathbf{p}}^{\zeta}-\gamma_{-\mathbf{k}}^{\zeta}-\gamma_{\mathbf{s}+\mathbf{r}-\mathbf{k}}^{\zeta}-\gamma_{\mathbf{s}}^{\zeta}\right)x_\mathbf{p}x_\mathbf{s}\nonumber\\
    &+\left(4\gamma_{\mathbf{r}-\mathbf{k}}^{\zeta}-\gamma_{\mathbf{r}-\mathbf{k}-\mathbf{p}}^{\zeta}-\gamma_{-\mathbf{k}}^{\zeta}-\gamma_{\mathbf{s}+\mathbf{r}-\mathbf{k}}^{\zeta}-\gamma_{\mathbf{r}}^{\zeta}\right)x_\mathbf{p}x_\mathbf{s}\nonumber\\
    &+\left(4\gamma_{\mathbf{p}-\mathbf{s}}^{\zeta}-\gamma_{\mathbf{p}-\mathbf{r}-\mathbf{s}}^{\zeta}-\gamma_{-\mathbf{s}}^{\zeta}-\gamma_{\mathbf{k}+\mathbf{p}-\mathbf{s}}^{\zeta}-\gamma_{\mathbf{p}}^{\zeta}\right)x_\mathbf{k}x_\mathbf{r}\nonumber\\
    &+\left(4\gamma_{\mathbf{k}-\mathbf{s}}^{\zeta}-\gamma_{\mathbf{k}-\mathbf{r}-\mathbf{s}}^{\zeta}-\gamma_{-\mathbf{s}}^{\zeta}-\gamma_{\mathbf{p}+\mathbf{k}-\mathbf{s}}^{\zeta}-\gamma_{\mathbf{k}}^{\zeta}\right)x_\mathbf{k}x_\mathbf{r}\nonumber\\
    &+\left(4\gamma_{\mathbf{p}-\mathbf{r}}^{\zeta}-\gamma_{\mathbf{p}-\mathbf{s}-\mathbf{r}}^{\zeta}-\gamma_{-\mathbf{r}}^{\zeta}-\gamma_{\mathbf{k}+\mathbf{p}-\mathbf{r}}^{\zeta}-\gamma_{\mathbf{p}}^{\zeta}\right)x_\mathbf{k}x_\mathbf{r}\nonumber\\
    &+\left(4\gamma_{\mathbf{k}-\mathbf{r}}^{\zeta}-\gamma_{\mathbf{k}-\mathbf{s}-\mathbf{r}}^{\zeta}-\gamma_{-\mathbf{r}}^{\zeta}-\gamma_{\mathbf{p}+\mathbf{k}-\mathbf{r}}^{\zeta}-\gamma_{\mathbf{k}}^{\zeta}\right)x_\mathbf{k}x_\mathbf{r},\\    
    \Phi_{\zeta}^{'}=&\Phi_{\zeta}^{''}=\left(4\gamma_{\mathbf{r}-\mathbf{p}}^{\zeta}-\gamma_{\mathbf{r}-\mathbf{k}-\mathbf{p}}^{\zeta}-\gamma_{-\mathbf{p}}^{\zeta}-\gamma_{\mathbf{s}+\mathbf{r}-\mathbf{p}}^{\zeta}-\gamma_{\mathbf{r}}^{\zeta}\right)\nonumber\\
    &+\left(4\gamma_{\mathbf{r}-\mathbf{p}}^{\zeta}-\gamma_{\mathbf{r}-\mathbf{k}-\mathbf{p}}^{\zeta}-\gamma_{-\mathbf{p}}^{\zeta}-\gamma_{\mathbf{s}+\mathbf{r}-\mathbf{p}}^{\zeta}-\gamma_{\mathbf{r}}^{\zeta}\right)x_\mathbf{k}x_\mathbf{p}x_\mathbf{s}x_\mathbf{r}.    
\end{align}

Collecting terms with the same four-magnon operator structure, we obtain the effective four-magnon scattering Hamiltonian:
\begin{align}
&H_{\rm scat}
=
H_1^{'}+H_2^{'}+H_3^{'}+H_4^{'} \nonumber\\
&=
-\dfrac{1}{N}
\sum_{\mathbf{k}\mathbf{p}\mathbf{s}\mathbf{r}}
\delta(\mathbf{k}+\mathbf{p}-\mathbf{s}-\mathbf{r}+G)\,
u_{\mathbf{k}}u_{\mathbf{p}}u_{\mathbf{s}}u_{\mathbf{r}}
\nonumber\\
&\times
\Bigg[
\alpha_{\mathbf{k}}^{\dagger}\alpha_{\mathbf{r}}
\beta_{\mathbf{s}}^{\dagger}\beta_{\mathbf{p}}
\left(
\dfrac{z_1J_1}{2}\Phi_1
+\dfrac{z_2J_2}{4}\Phi_2
+\dfrac{z_3J_3}{2}\Phi_3
+\dfrac{z_4J_4}{4}\Phi_4
\right)
\nonumber\\
&+
\alpha_{\mathbf{k}}^{\dagger}\alpha_{\mathbf{p}}^{\dagger}
\alpha_{\mathbf{s}}\alpha_{\mathbf{r}}
\left(
\dfrac{z_1J_1}{2}\Phi_1^{'}
+\dfrac{z_2J_2}{4}\Phi_2^{'}
+\dfrac{z_3J_3}{2}\Phi_3^{'}
+\dfrac{z_4J_4}{4}\Phi_4^{'}
\right)
\nonumber\\
&+
\beta_{\mathbf{k}}\beta_{\mathbf{p}}
\beta_{\mathbf{s}}^{\dagger}\beta_{\mathbf{r}}^{\dagger}
\left(
\dfrac{z_1J_1}{2}\Phi_1^{''}
+\dfrac{z_2J_2}{4}\Phi_2^{''}
+\dfrac{z_3J_3}{2}\Phi_3^{''}
+\dfrac{z_4J_4}{4}\Phi_4^{''}
\right)
\Bigg].
\label{eq:Hscat_app}
\end{align}

The matrix element $\mathcal M$ entering Fermi's golden rule is read from
Eq.~\eqref{eq:Hscat_app} as the coefficient of the corresponding
number-conserving four-magnon operator, excluding the overall factor $1/N$
and the momentum-conservation delta function. For example, for the
$\alpha\beta$ channel one has
\begin{eqnarray}
\mathcal M^{\alpha\beta;\alpha\beta}_{\mathbf{k}\mathbf{p};\mathbf{r}\mathbf{s}}&
=&-u_{\mathbf{k}}u_{\mathbf{p}}u_{\mathbf{s}}u_{\mathbf{r}}\nonumber\\
&\times& \left(
\dfrac{z_1J_1}{2}\Phi_1
+\dfrac{z_2J_2}{4}\Phi_2
+\dfrac{z_3J_3}{2}\Phi_3
+\dfrac{z_4J_4}{4}\Phi_4
\right),\nonumber\\
\end{eqnarray}
with analogous expressions for the $\alpha\alpha$ and $\beta\beta$ channels
obtained by replacing $\Phi_n$ with $\Phi_n^{'}$ and $\Phi_n^{''}$,
respectively.

\section{Zero Modes Associated with Conservation Laws}
\label{appC}

In this appendix, we derive the five zero modes of the linearized collision
operator when only momentum-conserving Normal scattering processes are
included. A zero mode is a distribution deviation that produces no relaxation
under the linearized collision operator. Equivalently, it can be obtained by
linearizing a family of stationary distributions of the collision integral.

We first recall that the Bose--Einstein distribution
\begin{equation}
f_{\mathbf{k}}^0
=
\frac{1}{\exp[\epsilon_{\mathbf{k}}/(k_BT)]-1}
\end{equation}
is an equilibrium solution of the collision integral in
Eq.~\eqref{eq:collision_integral}. Substituting $f_{\mathbf{k}}^0$ into the
gain and loss terms gives zero, as required by detailed balance. If a
modified Bose distribution is still stationary because of a conserved
quantity, then an infinitesimal variation within this family does not relax.
In the linearized theory, such a variation is therefore a zero mode of the
collision operator.

In the symmetrized representation used in this work, the distribution
deviation is written as
\begin{equation}
\delta f_{\mathbf{k}}
=
\sqrt{
f_{\mathbf{k}}^0
\left(
1+f_{\mathbf{k}}^0
\right)
}
\,
\phi_{\mathbf{k}} .
\label{eq:appC_sym_deviation}
\end{equation}
Thus, once the non-relaxing deviation $\delta f_{\mathbf{k}}$ is identified,
the corresponding zero mode is obtained by dividing by
$\sqrt{f_{\mathbf{k}}^0(1+f_{\mathbf{k}}^0)}$.

The energy zero mode is obtained by varying the temperature of the
equilibrium distribution. Since a temperature change produces another
equilibrium Bose distribution, this deviation cannot be relaxed by the
collision operator. Expanding $f_{\mathbf{k}}^0(T+\delta T)$ to linear order
gives
\begin{equation}
\delta f_{\mathbf{k}}
=
\frac{\partial f_{\mathbf{k}}^0}{\partial T}
\delta T
=
\frac{\epsilon_{\mathbf{k}}}{k_BT^2}
f_{\mathbf{k}}^0
\left(
1+f_{\mathbf{k}}^0
\right)
\delta T .
\end{equation}
Therefore, in the symmetrized representation, the energy zero mode is
\begin{equation}
\phi_E(\mathbf{k})
\propto
\epsilon_{\mathbf{k}}
\sqrt{
f_{\mathbf{k}}^0
\left(
1+f_{\mathbf{k}}^0
\right)
}.
\label{eq:appC_energy_zero_mode}
\end{equation}

The magnon-number zero mode can be derived in the same way by introducing a
chemical potential. Although the physical equilibrium distribution used in
the main text has $\mu=0$, the number-conserving four-magnon collision
operator also admits Bose distributions with finite $\mu$ as stationary
solutions. We therefore consider
\begin{equation}
f_{\mathbf{k}}^0(\mu)
=
\frac{1}
{\exp[(\epsilon_{\mathbf{k}}-\mu)/(k_BT)]-1}.
\end{equation}
Expanding around $\mu=0$ gives
\begin{equation}
\delta f_{\mathbf{k}}
=
\left.
\frac{\partial f_{\mathbf{k}}^0}{\partial \mu}
\right|_{\mu=0}
\delta\mu
=
\frac{1}{k_BT}
f_{\mathbf{k}}^0
\left(
1+f_{\mathbf{k}}^0
\right)
\delta\mu .
\end{equation}
The corresponding zero mode is therefore
\begin{equation}
\phi_N(\mathbf{k})
\propto
\sqrt{
f_{\mathbf{k}}^0
\left(
1+f_{\mathbf{k}}^0
\right)
}.
\label{eq:appC_number_zero_mode}
\end{equation}

The remaining three zero modes are associated with crystal-momentum
conservation in Normal scattering. For Normal processes, the total crystal
momentum is conserved, and the collision integral is also stationary for a
drifted Bose distribution,
\begin{equation}
f_{\mathbf{k}}^{(\mathbf{u})}
=
\frac{1}
{\exp[(\epsilon_{\mathbf{k}}-\mathbf{u}\cdot\mathbf{k})/(k_BT)]-1},
\end{equation}
where $\mathbf{u}$ is a small drift velocity. Expanding this distribution
with respect to $u_a$ gives
\begin{eqnarray}
\delta f_a(\mathbf{k})
&=&
\left.
\frac{\partial f_{\mathbf{k}}^{(\mathbf{u})}}{\partial u_a}
\right|_{\mathbf{u}=0}
\delta u_a\nonumber\\
&=&
\frac{k_a}{k_BT}
f_{\mathbf{k}}^0
\left(
1+f_{\mathbf{k}}^0
\right)
\delta u_a,
\quad
a=x,y,z .
\end{eqnarray}
Thus the three momentum zero modes are
\begin{equation}
\phi_{P_a}(\mathbf{k})
\propto
k_a
\sqrt{
f_{\mathbf{k}}^0
\left(
1+f_{\mathbf{k}}^0
\right)
},
\qquad
a=x,y,z .
\label{eq:appC_momentum_zero_mode}
\end{equation}

Together, the five zero modes of the Normal-scattering collision operator are
\begin{equation}
\phi_N(\mathbf{k})
\propto
\sqrt{
f_{\mathbf{k}}^0
\left(
1+f_{\mathbf{k}}^0
\right)
},
\qquad
\phi_E(\mathbf{k})
\propto
\epsilon_{\mathbf{k}}
\sqrt{
f_{\mathbf{k}}^0
\left(
1+f_{\mathbf{k}}^0
\right)
},
\end{equation}
and
\begin{equation}
\phi_{P_a}(\mathbf{k})
\propto
k_a
\sqrt{
f_{\mathbf{k}}^0
\left(
1+f_{\mathbf{k}}^0
\right)
},
\qquad
a=x,y,z .
\end{equation}
They correspond respectively to magnon-number conservation, energy
conservation, and the three components of crystal-momentum conservation.

\section{Relation between the momentum modes and the heat-current driving vector}
\label{appD}

In this appendix, we explain why the momentum-related modes discussed in
App.~\ref{appC} have a large overlap with the heat-current driving vector
appearing in the LBTE. In the symmetrized formulation used in this work, the driving vector for a temperature gradient along direction $a$ is, up to an overall prefactor,
\begin{equation}
X_{\mathbf{k},a}
=
\epsilon_{\mathbf{k}}v_{\mathbf{k},a}
\sqrt{
f_{\mathbf{k}}^{0}
\left(1+f_{\mathbf{k}}^{0}\right)
},
\label{eq:appD_X}
\end{equation}
where $v_{\mathbf{k},a}$ is the magnon group velocity. On the other hand, the momentum zero mode of the
Normal-scattering collision operator has the form
\begin{equation}
\psi_{P_a,\mathbf{k}}
\propto
k_a
\sqrt{
f_{\mathbf{k}}^{0}
\left(1+f_{\mathbf{k}}^{0}\right)
},
\qquad
a=x,y,z .
\label{eq:appD_momentum_mode}
\end{equation}

For the low-energy $\epsilon_{\mathbf{k}}\lesssim
20\,{\rm meV}$, corresponding to about $0.59$ of the full magnon bandwidth, the spin-wave dispersion can be well approximated by
\begin{equation}
\epsilon_{\mathbf{k}}
\simeq
\sqrt{
c_{\perp}^{2}(k_x^2+k_y^2)
+
c_z^2 k_z^2+c^2_0
}.
\label{eq:appD_low_energy_dispersion}
\end{equation}
For the spin Hamiltonian used in this work,
\begin{eqnarray}
&c^2_{\perp}
=2 a^2 S^2 [J_3z_3 (J_1 z_1 + J_3 z_3) - J_2z_2 (K + J_1 z_1 + J_3 z_3)],\nonumber\\
\\
&c^2_z=c^2S^2 (J_1 z_1 + J_3 z_3)^2 - 
 4 c^2 J_4 S^2z_4 (K + J_1 z_1 + J_3 z_3),\nonumber\\
\end{eqnarray}
and
\begin{equation}
c^2_0=4 (K^2 S^2 + 2 J_1 K S^2 z_1 + 2 J_3 K S^2 z_3).
\end{equation}

Using Eq.~\eqref{eq:appD_low_energy_dispersion}, the group-velocity
components are
\begin{equation}
v_{\mathbf{k},x}
=
\frac{\partial \epsilon_{\mathbf{k}}}{\partial k_x}
=
\frac{c_{\perp}^{2}k_x}{\epsilon_{\mathbf{k}}},
\qquad
v_{\mathbf{k},y}
=
\frac{c_{\perp}^{2}k_y}{\epsilon_{\mathbf{k}}},
\qquad
v_{\mathbf{k},z}
=
\frac{c_z^{2}k_z}{\epsilon_{\mathbf{k}}}.
\end{equation}
Therefore,
\begin{equation}
\epsilon_{\mathbf{k}}v_{\mathbf{k},a}
=
c_a^2 k_a,
\qquad
c_x=c_y=c_{\perp},
\quad
c_z=c_z .
\label{eq:appD_epsilon_v}
\end{equation}
Substituting this relation into Eq.~\eqref{eq:appD_X} gives
\begin{equation}
X_{\mathbf{k},a}
\propto
k_a
\sqrt{
f_{\mathbf{k}}^{0}
\left(1+f_{\mathbf{k}}^{0}\right)
}
\propto
\psi_{P_a,\mathbf{k}}.
\label{eq:appD_X_momentum}
\end{equation}

This result shows that, in the low-energy regime, the heat-current driving
vector has the same momentum-space structure as the corresponding momentum
zero modes of the Normal-scattering collision operator. Therefore, the thermal
driving field couples efficiently to these modes. When Umklapp
scattering is included, the exact momentum zero modes are lifted to finite
relaxation rates; however, the resulting low-rate eigenmodes retain momentum-like
character and can acquire large weights in the mode-resolved decomposition of
the thermal conductivity.
\bibliography{ref}

\end{document}